\documentstyle[12pt]{article}

\def\Int{{\int}\!\!\mathop{\vphantom{\int}}}
\def\Lvec#1{\overleftarrow{#1}}

\def\di{\partial}

  \def\pp{{\mathchoice
            {
              \kern 1pt%
              \raise 1pt
              \vbox{\hrule width5pt height0.4pt depth0pt
                    \kern -2pt
                    \hbox{\kern 2.3pt
                          \vrule width0.4pt height6pt depth0pt
                          }
                    \kern -2pt
                    \hrule width5pt height0.4pt depth0pt}%
                    \kern 1pt
           }
            {
              \kern 1pt%
              \raise 1pt
              \vbox{\hrule width4.3pt height0.4pt depth0pt
                    \kern -1.8pt
                    \hbox{\kern 1.95pt
                          \vrule width0.4pt height5.4pt depth0pt
                          }
                    \kern -1.8pt
                    \hrule width4.3pt height0.4pt depth0pt}%
                    \kern 1pt
            }
            {
              \kern 1pt%
              \raise 1pt
              \vbox{\hrule width3.8pt height0.4pt depth0pt
                    \kern -1.6pt
                    \hbox{\kern 1.7pt
                          \vrule width0.4pt height4.8pt depth0pt
                          }
                    \kern -1.6pt
                    \hrule width3.8pt height0.4pt depth0pt}%
                    \kern 1pt
            }
            {
              \kern 1pt%
              \raise 1pt
              \vbox{\hrule width3.6pt height0.35pt depth0pt
                    \kern -1.5pt
                    \hbox{\kern 1.625pt
                          \vrule width0.4pt height4.5pt depth0pt
                          }
                    \kern -1.5pt
                    \hrule width3.6pt height0.35pt depth0pt}%
                    \kern 1pt
            }
        }}

  \def\mm{{\mathchoice{
                             \kern 1pt
               \raise 1pt    \vbox{\hrule width5pt height0.4pt depth0pt
                                  \kern 2pt
                                  \hrule width5pt height0.4pt depth0pt}
                             \kern 1.5pt}
                       {
                            \kern 1pt
               \raise 1pt \vbox{\hrule width4.3pt height0.4pt depth0pt
                                  \kern 1.8pt
                                  \hrule width4.3pt height0.4pt depth0pt}
                             \kern 1pt}
                       {
                            \kern 1pt
               \raise 1pt
                            \vbox{\hrule width3.8pt height0.4pt depth0pt
                                  \kern 1.6pt
                                  \hrule width3.8pt height0.4pt depth0pt}
                            \kern 1pt}
                       {
                           \kern 1pt
             \raise 1pt  \vbox{\hrule width3.6pt height0.35pt depth0pt
                                  \kern 1.5pt
                                  \hrule width3.6pt height0.35pt depth0pt}
                           \kern 1pt}
                       }}

\def\diz{\partial_{\pp}^{z}}
\def\dizp{\partial_{\pp}^{z'}}
\def\Dz{D_{{\scriptscriptstyle +}}^{z}}
\def\Dzp{D_{+}^{z'}}

\def\lp{\lambda^{+}}

\def\S#1{\frac{1}{{S^{\pp}}^{#1}}}

\def\downprop{\rule[-8pt]{0pt}{0pt}}

    \def\tic{\vrule height0.4pt width0.4pt depth2pt}
  \def\mAth{\mathsurround=0pt }
  \def\contractionfill#1#2{$\mAth
                \setbox0=\hbox{$\displaystyle{#1}$}\kern0.5\wd0\kern-0.2pt
                \rlap{\tic}\leaders\hrule\hfil\llap{\tic}
                \setbox0=\hbox{$\displaystyle{#2}$}\kern0.5\wd0\kern-0.2pt$}
  \def\contract#1#2#3{\mathop{\vbox{\ialign{##\crcr\noalign{\kern3pt}
        \contractionfill{#1}{#3}\crcr\noalign{\kern3pt\nointerlineskip}
        $\hfil\displaystyle{#1#2#3}\hfil$\crcr}}}\limits}

\renewcommand{\thesubsection}{\arabic{subsection}}

\def\beq{\begin{equation}}
\def\eeq{\end{equation}}
\def\bea{\begin{eqnarray}}
\def\eea{\end{eqnarray}}
\def\bq{\begin{quote}}
\def\eq{\end{quote}}

\renewcommand{\thefootnote}{\fnsymbol{footnote}}

\begin{document}

\begin{titlepage}
\begin{flushright}
{BRX-TH-361}
\end{flushright}
\vspace{2cm}
\begin{center}
{\bf {\large   OPE's and the Dilaton Beta-Function for the\\
                     2-D N=1 Supersymmetric Non-Linear $\sigma$-Model}}\\
\vspace{1.5cm}
Marcia E. Wehlau \footnote{\hbox to \hsize{Current address:
Mars Scientific Consulting, 28 Limeridge Dr., Kingston, ON CANADA K7K~6M3}}\\
\vspace{1mm}
{\em Physics Department, Brandeis University, Waltham, MA 02254, USA}\\
\vspace{1.1cm}
{ABSTRACT}
\end{center}

\bq
Using the superspace formalism, we compute for the
two-dimensional N=1 supersymmetric non-linear $\sigma$-model, the order
 $(\alpha^{\prime})^{2}$
$(R_{mnpq})^2$  (three-loop) correction to the central charge via the
operator
product expansion of the supercurrent with itself. The contribution vanishes,
in
agreement with previous results obtained from the usual $\sigma$-model
$\beta$-function approach.
\eq

\vfill

\begin{flushleft}
\today
\end{flushleft}
\end{titlepage}

\renewcommand{\thefootnote}{\arabic{footnote}}
\setcounter{footnote}{0}
\pagenumbering{arabic}

\subsection{Introduction}\label{Introduction}

Two-dimensional superconformal field theory (SCFT) has proven to be
a useful means for investigating the relationship between supersymmetric
non-linear sigma-models and superstrings in background fields.
Conformal invariance is necessary for consistent string propagation
in a curved background, and this effectively forces the $\beta$-functions
 of the corresponding $\sigma$-model
 to be zero.  The equations $\beta=0$ so obtained can then be
  identified as the equations of motion of the background fields ---
the metric $G_{ij}(X)$,
the antisymmetric tensor $B_{ij}(X)$, and the dilaton $\Phi(X)$.
  The results of string computations
  have been found to be in agreement with those from $\sigma$-model
  calculations \cite{Frad,CFMP,Brust}.

 From SCFT we get expressions for the
operator product expansion (OPE) of two operators. It is possible to use
these OPE's as an alternative means of obtaining the $\sigma$-model
$\beta$-functions, instead of using the standard
renormalization group procedures.  Specifically,
the expectation value $<J(z)\,J(z')>$, where
  the operator $J$ is the supercurrent, can be computed perturbatively for
the $\sigma$-model.  Once computed, $<J(z)\,J(z')>$ can be compared with
 the result
  of $J(z)J(z')$ from the operator product expansion.  Calculations of this
  type have been done for the bosonic $\sigma$-model
  \cite{BNS}, and for the N=1 supersymmetric case (to one-loop order for
$\beta_{ij}^G$ and $\beta_{ij}^B$, and to two-loop order for $\beta^\Phi$)
  \cite{AHZ}.
  In these calculations, additional terms appear in the perturbation
expansion  of  $<J(z)\,J(z')>$ that do not exist in the OPE for
$J(z)J(z')$.  These extra terms can be observed to be
essentially the $\beta$-functions of the $\sigma$-model
and when set to zero for consistency with the OPE's, yield the background
equations of motion for the superstring fields.

   In this paper we apply this
alternate method of computing the $\beta$-functions to the N=1 supersymmetric
$\sigma$-model (for the case $B_{ij} = 0$), by
examining $<J(z)\,J(z')>$ at three loops for extra contributions to the
central charge of the
form $(R_{ijkl})^{2}$, where $R_{ijkl}$ is the background Riemann tensor.
Terms of this form appear at two loops for the
bosonic $\sigma$-model
when the metric $\beta$-function is computed using the usual approach,
but they do not appear in the supersymmetric case
\cite{Gaume,Gaume2,GVdVZ,Allen,Ketov2}.
This makes the OPE calculation of particular interest.  We would like
to use the OPE method to obtain $\beta_{ij}^G$ directly, and thus determine
whether or not there are any new two-loop corrections.
However, corrections to the central charge are given by $\beta^\Phi$, and
from the usual $\sigma$-model $\beta$-function results,
$\beta_{ij}^{G(L)} \sim
 \frac {\textstyle \delta\beta^{\Phi(L+1)}} {\textstyle \delta G_{ij}}\ ,$
where $L$ is the number of loops.  Because the OPE generates quite a large
number of two-loop diagrams that contribute to $\beta_{ij}^G$, we compute
instead contributions to $\beta^\Phi$ at three loops (from a considerably
smaller set of diagrams), of the form $(R_{mnpq})^2$.  These terms are the
ones which could lead to new contributions to $\beta_{ij}^G$ of the form
${R_i}^{klm} R_{jklm}$.  From previous calculations
\cite{Gaume,Gaume2,GVdVZ,Allen,Ketov2}, the contribution from the extra
$(R_{ijkl})^{2}$ terms is expected to be zero and the result of our
calculation indicates that this is indeed the case.

The paper is organized as follows: In section 2, we give the action of the
N =1 supersymmetric non-linear $\sigma$-model, the OPE for the supercurrent
with itself, and the background field expansions for both of them.
Section 3 discusses the calculation of the three-loop correction to the
central charge.  Our notation and conventions are those of
\cite{GGRS} and are listed in the appendix, along with a discussion of the
techniques used and a sample diagram computation.  Details of the calculation
can be found in ref. \cite{thesis}.

  \setcounter{equation}{0}
  \subsection{Action, OPE and the Background Field Expansion}

 The use of OPE's for obtaining the supersymmetric $\sigma$-model
 $\beta$-functions was first presented in \cite{AHZ}.
{}From superconformal field theory, the supercurrent ${J_+}^\mm$
must satisfy the OPE
\begin{equation}
    {J_+}^\mm(z) {J_+}^\mm(z') \sim  \frac{c/4}{{S^\pp}^3}
  +               \frac{3} {2}\frac{\lambda^+}{{S^\pp}^2} {J_+}^\mm(\Sigma) +
   \frac{1}{2}    \frac{D_+ {J_+}^\mm(\Sigma)}{S^\pp} +{\rm\ finite\ terms}\ ,
    \label{JJ OPE}
\end{equation}
  where $S^\pp = x^\pp - x'^\pp -i \theta^+ {\theta'}^+$ is the supersymmetric
  coordinate difference, $\lambda^\pm = {\scriptstyle\frac{1}{2}}
(\theta~-~{\theta'})^\pm$, the midpoint is
 $\Sigma = \left\{{\scriptstyle \frac{1}{2}}{(x + x')^\pm},\,
{\scriptstyle \frac{1}{2}}{(\theta + {\theta'})^\pm}\right\}$, and $c$ is the
central charge.
  The procedure we follow now is just as in the previous bosonic \cite{BNS}
 and supersymmetric cases \cite{AHZ}  ---  calculate the expectation value
  $<{J_+}^\mm(z)\, {J_+}^\mm(z')>$, and demand that (\ref{JJ OPE}) hold true.

We start with the action for the non-linear $\sigma$-model coupled to 2-D
supergravity
   \begin{equation}
   S = -\frac{1}{4\pi\alpha'}\Int d^2x\, d^2\theta E^{-1} [ G_{ij}(X)
          {\cal D}_+ X^i {\cal D}_- X^j  + \frac{\alpha'}{2} R^{(2)} \Phi(X) ]
             \ ,
           \label{action}
    \end{equation}
where $\cal D_{\pm}$ are the supergravity covariant derivatives,
the superfield $R^{(2)}$ is the scalar curvature of the 2-D worldsheet,
and $E^{-1}$ is the superdeterminant of the inverse superzweibein.
The supercurrent ${J_+}^\mm$ is defined to be:
\begin{equation}
   {J_+}^\mm = \frac{2\delta S}{\delta {H_+}^\mm}{\biggm |}_{{H_+}^\mm = 0}\ .
\end{equation}
  ${H_+}^\mm$ is the supergravity gauge field, and the
  supercurrent is the variation of the action with respect to this field.
We find
\begin{equation}
      {J_+}^\mm = -\frac{1}{2\pi\alpha'}[ G_{ij}(X) \di_\pp X^i D_+ X^j
             - \alpha' \di_\pp D_+ \Phi(X) ] \   .
\end{equation}

We use the background field method \cite{AM} as it applies to superfields
to perform the perturbative expansion of $<{J_+}^\mm(z) {J_+}^\mm(z')>$ .
This involves expanding the action and the supercurrent in terms of Riemann
normal coordinates $\xi^i$ on the background manifold. The
relevant expansions are:
\newpage
\begin{eqnarray}
  \di_\pp X^i & = & \di_\pp X^i_B + \nabla_\pp\xi^i +
    {\scriptstyle \frac 1 3} R^i_{lmn}(X_B) \di_\pp X^n_B \xi^l\xi^m +
{\scriptstyle \frac 1 {12}}D_j {R^i}_{lmn}(X_B)\xi^j\xi^l\xi^m\di_\pp X^n_B
   \nonumber\\
&& +{\scriptstyle\frac 1 {60}}D_j D_k {R^i}_{lmn}(X_B)
               \di_\pp X^n_B \xi^j\xi^k\xi^l\xi^m
   -{\scriptstyle\frac 1 {45}}{R^i}_{jkp}{R^p}_{lmn}(X_B)
             \di_\pp X^n_B \xi^j\xi^k\xi^l\xi^m + \ldots\downprop \nonumber\\
D_+X^j &=& D_+X^i_B + \nabla_+\xi^i +
  {\scriptstyle \frac 1 3}{R^i}_{lmn}(X_B) D_+ X^n_B \xi^l\xi^m
  + {\scriptstyle \frac 1 {12}}D_j {R^i}_{lmn}(X_B) D_+ X^n_B \xi^j\xi^l\xi^m
\downprop  \nonumber\\
&& + {\scriptstyle\frac 1 {60}} D_j D_k {R^i}_{lmn}(X_B)
           D_+ X^n_B \xi^j\xi^k\xi^l\xi^m
   - {\scriptstyle\frac 1 {45}}{R^i}_{jkp}{R^p}_{lmn}(X_B)
           D_+ X^n_B \xi^j\xi^k\xi^l\xi^m  + \ldots \downprop\nonumber\\
  G_{ij}(X) & = & G_{ij}(X_B) - {\scriptstyle \frac 1 3}
       R_{ikjl}(X_B)\xi^k\xi^l
    -{\scriptstyle \frac 1 {3!}} D_l R_{imjk}(X_B)\xi^l \xi^m \xi^k \\
&& + {\scriptstyle \frac 1 {5!}}\left(-6D_k D_l R_{imjn}(X_B) +
     {\scriptstyle \frac {16} 3}{R_{kjl}}^p R_{minp}(X_B)\right)
      \xi^k \xi^l \xi^m \xi^n + \ldots \downprop \nonumber\\
  \Phi(X) &=& \Phi(X_B) + D_i\Phi(X_B)\xi^i + {\scriptstyle \frac 1 2}
     D_iD_j\Phi(X_B)\xi^i\xi^j + \ldots \nonumber
\end{eqnarray}
  where the background covariant derivatives are
  \[
   \nabla_\pp\xi^i = \di_\pp\xi^i + \Gamma^i_{jk}(X_B)\xi^i\di_\pp X_B^k\
{\rm\ and\ }
   \nabla_+\xi^i = D_+\xi^i + \Gamma^i_{jk}(X_B) \xi^j D_+ X_B^k\ ,
  \]
and $X_B$ is the background field.
The expansions of the action and the supercurrent (where the subscript
$B$ has been dropped) are:
\begin{eqnarray}
S &=& \frac 1 {8\pi\alpha'} \Int d^2x\, d^2\theta \bigg\{
  G_{ij}(X) D_\alpha X^i D^\alpha X^j \
    +\ 2G_{ij}D_\alpha X^i \nabla^\alpha \xi^j \nonumber\\
&&  \ +\ G_{ij}\nabla_\alpha\xi^i \nabla^\alpha\xi^j
  +\ R_{ijkl}D_\alpha X^i D^\alpha X^j \xi^k \xi^l \label{L AHZ}\\
&&\ +\ {\scriptstyle\frac 4 3}R_{ijkl}D_\alpha X^i\nabla^\alpha\xi^j\xi^k\xi^l
    +\ {\scriptstyle\frac 1 3}
           R_{ijkl} \nabla_\alpha \xi^i \nabla^\alpha \xi^l \xi^j \xi^k
    \ +\ \ldots \bigg\} \ .  \nonumber\\
J_+{}^\mm & = & -\frac{1}{2\pi\alpha'} \biggl\{ G_{ij} \di_\pp X^i D_+ X^j
 + G_{ij}D_+ X^i \nabla_\pp \xi^j\ +\ G_{ij}\di_\pp X^i\nabla_+\xi^i
   \nonumber \\
 &&+\ G_{ij}\nabla_+ \xi^i \nabla_\pp \xi^j
       \ +\ R_{ilmj}\xi^l\xi^m D_+ X^i \di_\pp X^j \
+\ {\scriptstyle\frac 1 3}R_{ijkl}\nabla_+\xi^i \nabla_\pp\xi^l \xi^k \xi^k\
\nonumber\\
&&+\ {\scriptstyle \frac 2 3} R_{ilmn}\di_\pp X^n \nabla_+ \xi^i \xi^l \xi^m\
 +\ {\scriptstyle \frac 2 3} R_{ilmn} D_+ X^n \nabla_\pp \xi^i \xi^l \xi^m
                           \label{J AHZ}\\
&& -\ \alpha' \Bigl[ D_iD_j\Phi D_+ X^i \nabla_\pp \xi^j
       \,+\,D_i D_j\Phi\di_\pp X^j \nabla_+ \xi^j \nonumber\\
&&\qquad +\, D_i\Phi\nabla_\pp \nabla_+ \xi^i \,+\,
  {\scriptstyle \frac 1 2} D_i D_j\Phi \nabla_\pp\nabla_+ (\xi^i \xi^j)
    \,+\, \ldots \Bigr]
    \biggr\}\ . \nonumber
\end{eqnarray}

  Following the standard procedure, we refer $\xi$ to the tangent frames
on the manifold,  $\xi^a = {E^a}_i \xi^i$ , so that $\xi^a$ is the quantum
field used in the calculation and
 \mbox{$\nabla_\alpha \xi^a = D_\alpha \xi^a + {\omega_i}^a_b D_\alpha X^i
  \xi^b$}, where ${\omega_i}^a_b$ is the spin connection.

  \setcounter{equation}{0}
  \subsection{The Three-Loop Correction to the Central Charge} \label{three
loop}

 We
  calculate the order $(\alpha')^2 (R_{abcd})^2$ contributions to the central
  charge, and these involve three-loop graphs.  {\it A priori\/} we expect the
 result to be zero because as mentioned previously, we know that
 the usual method of calculating $\beta_{ij}^G$ indicates that it
  receives no two-loop corrections in the supersymmetric case.

It turns out that the expansions given previously in section 2 are
sufficient to generate the four different types of graphs that yield
  $(R_{abcd})^2$ contributions.  We isolate below the relevant interaction
  vertices from the Lagrangian and the supercurrent that make up these graphs:

  Lagrangian term:
\begin{eqnarray}
  {{\cal L}_{int}}  &=& {\textstyle \frac 1 3} R_{dfge} D_+ \xi^d D_-
                      \xi^e\xi^f\xi^g
       \label{L term}
\end{eqnarray}

Supercurrent terms:
\begin{eqnarray}
J &=& G_{cd} \di_\pp \xi^c D_+ \xi^d
       + {\textstyle \frac 1 3}R_{abcd}\xi^c \xi^d \di_\pp \xi^b D_+ \xi^a
   \label{J terms}
\end{eqnarray}
  The diagrams are evaluated entirely in coordinate-space using propagators
$G(z, z')$.
  We recall that a conserved quantity such as the supercurrent
  should not receive any renormalization counterterms.  In the
  one-loop case \cite{BNS,AHZ}, all the divergences that did not cancel out
  between
  diagrams could be isolated into a tadpole integral, $G(0)$. This $G(0)$
  divergence cancelled out of both sides of (\ref{JJ OPE}) for the
  supercurrent because, although it appeared in $<JJ>$ on the left-hand side
  of the OPE, it also appeared in $<J>$ on the right.  Because we expect that
  ultimately the supercurrent should be finite, we evaluated the graphs in such
  a way that we kept only the finite pieces, while noting that all
  the divergent terms we discard are of the form $G(0)$, or $\delta(0) G(0)$.
  However, we did not explicitly check this, and likewise we did not
 explicitly show that contributions from the connection terms cancel
 among themselves.  The
  latter was verified in the bosonic and one-loop supersymmetric cases, and
  is expected to be generally true.  We used standard superspace techniques
  \cite{GGRS} in the actual calculation of the diagrams, paying particular
  attention when
  doing the $D$-algebra to the fact that one is {\em not} computing an
  effective action here, but an OPE.  Because of the structure of the OPE,
  the points $z$ and $z'$ are not integrated over, and this means that one
  cannot push $D$'s past those points on the diagrams.

    The superspace propagator in coordinate-space is
\begin{equation}
\Delta(z,z') = \contract{\xi^i}{(z)}{\xi^j}(z')
                     = -\frac{1}{2} \alpha' \delta^{ij} \ln(S^\pp S^\mm)
                       = -\frac 1 2 \alpha' \delta^{ij} G(z,z')
   \label{propagator}
\end{equation}
  and in doing the D-algebra we eliminate propagators from the diagrams (thus
  bringing two superspace points $u$ and $v$ together) by applying the
  following equation:
\begin{equation}
  D_-^u D_+^u \Delta(u,v) = -2\pi \alpha' \delta^{(4)}(u-v)
  \label{delta}
\end{equation}
where $\Delta(u,v)$ is understood to be suitably defined by a cut-off $\mu$
whenever this equation is used.

  The four types of graphs that contribute to the $(R_{abcd})^2$ term in the
central
  charge are shown in Figure~1.  All the graphs yield contributions of the
  general form
\begin{equation}
   \S{3} R_{i(jk)l} R^{ijkl} \left[ G(z,z') + G(z,z')^2\right]
\end{equation}
  where $G(z,z') = \ln (S^\pp S^\mm) = \ln S^\pp + \ln S^\mm$, formally.

\let\picnaturalsize=N
\def\picsize{4.0in}
\def\picfilename{fig1.eps}
\ifx\nopictures Y\else{\ifx\epsfloaded Y\else\input epsf \fi
\let\epsfloaded=Y
\centerline{\ifx\picnaturalsize N\epsfxsize \picsize\fi
\epsfbox{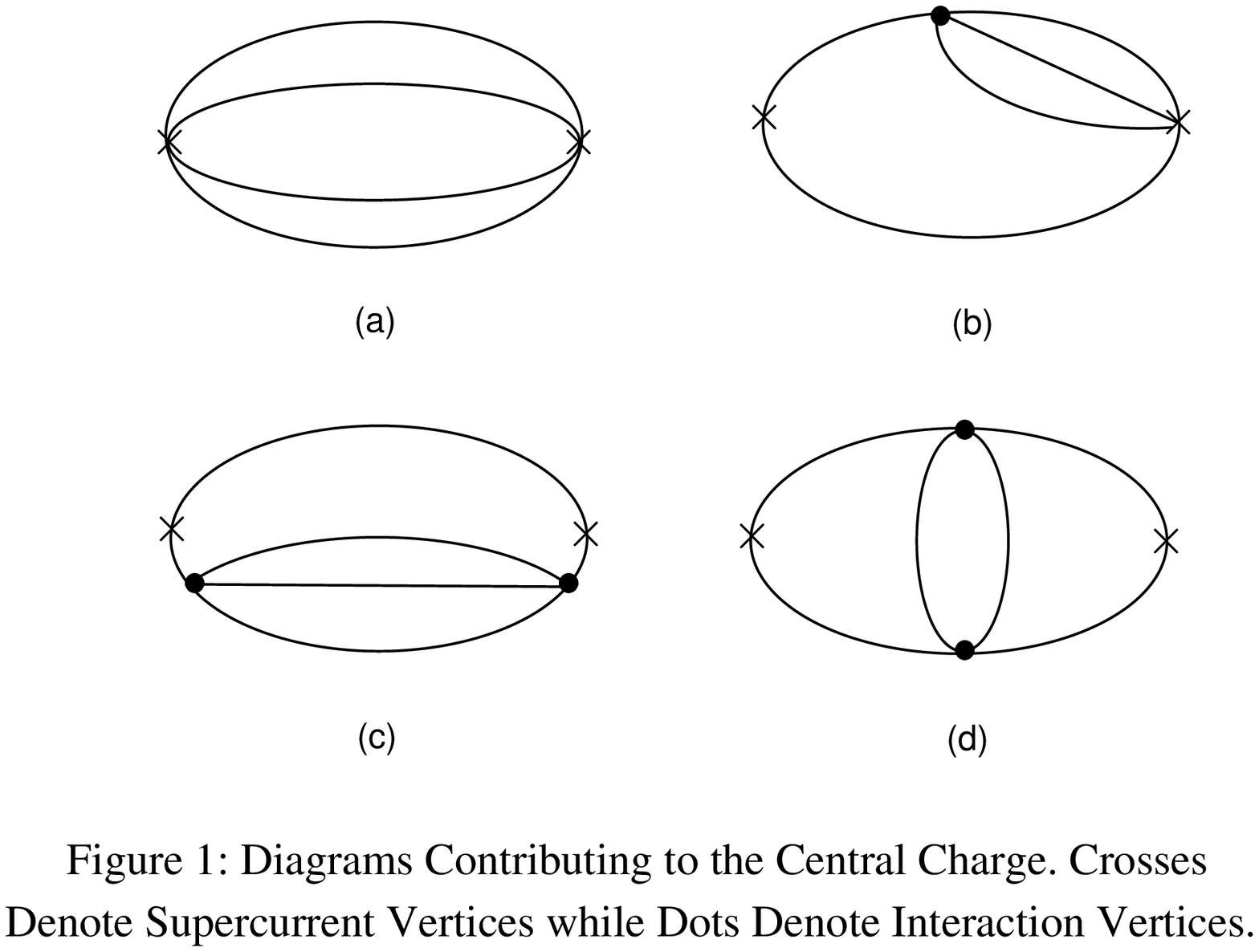}}}\fi
\vspace{5 mm}

  The perturbation expansion of $<{J_+}^\mm(z) {J_+}^\mm(z')>$  generates
 large numbers of specific
contributions for the four types of graphs shown in Figure~1.
  Initially there are over 300 different permutations of the derivatives for
  the four graphs, but many of these can be shown to be zero by D-algebra,
 or because
  $\lambda^+ \lambda^+ = 0$, or are discarded because they give contributions
involving $G(0)$.
 Tables of the relevant diagrams and their associated
 contributions can be found in \cite{thesis}.
  A sample calculation of one of the graphs of type (d) is given in the
 appendix.

Combining all the separate contributions for the various types of graphs
gives:\\
Diagrams of type (a):
\begin{equation}
\frac 1 9 \frac{(\alpha')^2} {256\pi^2}\frac i{{S^\pp}^3} R_{a(bc)d} R^{abcd}
   \left[ 4G(z,z') + 4G(z,z')^2 \right]
  \label{diagram a}
\end{equation}
Diagrams of type (b):
\begin{equation}
\frac 1 9\frac{(\alpha')^2} {256\pi^2}\frac i {{S^\pp}^3} R_{a(bc)d} R^{abcd}
   \left[ -8G(z,z') - 8G(z,z')^2 \right]
\label{diagram b}
\end{equation}
Diagrams of type (c):
\begin{equation}
\frac 1 9\frac{(\alpha')^2} {256\pi^2}\frac i {{S^\pp}^3} R_{a(bc)d} R^{abcd}
   \left[ -8G(z,z') + 8G(z,z')^2 \right]
  \label{diagram c}
\end{equation}
Diagrams of type (d):
\begin{equation}
\frac 1 9\frac{(\alpha')^2} {256\pi^2} \frac i {{S^\pp}^3} R_{a(bc)d} R^{abcd}
   \left[ 12G(z,z') - 4G(z,z')^2 \right]
 \label{diagram d}
\end{equation}

    By summing the above four terms, we obtain the complete result, which is
zero as expected.  Therefore we find that there are no additional corrections
to
$\beta_{ij}^G$ of the form ${R_i}^{klm} R_{jklm}$ because there are no
$(R_{ijkl})^2$ corrections to $\beta^\Phi$.

  \setcounter{equation}{0}
  \subsection{Conclusion}

  We have used results from superconformal field theory involving operator
 product expansions to obtain information about the $\beta$-functions of the
 N=1 supersymmetric non-linear $\sigma$-model. In particular, by
  computing the operator product expansion of the supercurrent with itself,
 we identified terms of the form $(\alpha')^2 (R_{ijkl})^2$ that might
  contribute to the central charge,
and hence give a new contribution to the dilaton $\beta$-function at
 three-loops. However, when all the terms are summed,  the
  complete result is found to be zero, in agreement with results obtained
 previously using  standard methods.

\def\thesubsection{} 
\subsection{Acknowledgements}

We would like to thank E.K. for amicably spirited discussions, and the Physics
Departments of the University of Toronto and Queen's University for their
hospitality. This work was supported in part by a NSERC Postgraduate
Scholarship.

\subsection{Appendix}

  (1,1) superspace is four-dimensional space that is parametrized by two
  commuting coordinates $x^a\ (a=0,1)$, and two anticommuting coordinates
  $\theta^\alpha\ (\alpha=1,2)$.   We use a
  one-component spinor notation ${\scriptstyle (+,-)}$ in which
  $\scriptstyle +\ (-)$ corresponds to the $+1/2\ (-1/2)$ helicity
  representation.  The spinor coordinates are denoted by
  $\theta^+$ and $\theta^-$, and the light-cone components of $x^a$ are

  \[ x^{\pp}= \frac{1}{\sqrt{2}}(x^0 + x^1) {\rm \ and\ }
     x^{\mm}= - \frac{1}{\sqrt{2}}(x^0 - x^1) \]

Our conventions
are those of \cite{GGRS}, but in $\scriptstyle (+,-)$ notation,
the superspace derivatives become
  \[D_A  = (\di_\pp,\di_\mm,D_+,D_-)\]
  where
  \[D_+ = \frac{\di}{\di\theta^+} + i\theta^+\di_\pp {\rm\ and \ }
    D_- = \frac{\di}{\di\theta^-} + i\theta^-\di_\mm  \   .\]

We use the expressions listed below, which can be obtained from those in
\cite{GGRS}
by carefully replacing a spinor index $\alpha$ with $+$ or $-$, and a
vector index $a$ (or a pair of spinor indices $\alpha\beta$) by $\pp$ or
$\mm$.

   \[\begin{array}{c}
    D_\alpha = \varepsilon_{\alpha\beta}D^\beta \ \rightarrow\ D_+ =
          \varepsilon_{-+}D^- = -D^- {\rm\ and\ } D_- = \varepsilon_{-+}D^+ =
            D^+  \\
      D^2 = \frac{1}{2}D^\alpha D_\alpha = \frac{1}{2}(D^+D_+ + D^-D_-) = D_-
               D_+ = D^-D^+ \\
      \{D_\alpha, D_\beta\} = 2i\di_{\alpha\beta} \ \rightarrow\ \{D_+, D_-\}=
            0,\{D_+, D_+\} = 2i\di_\pp, \{D_-,D_-\} = 2i\di_\mm  \\
      {(D^2)}^2 = \Box \quad \Box = \di_\mm\di_\pp = -p^2 {\rm\ if\ } i\di
         \ \rightarrow\ p \\
      \theta^2 = \frac{1}{2} \theta^\alpha\theta_\alpha = \theta_-\theta_+ =
         \theta^-\theta^+ \\
      D^2D_+ = -D_+D^2 = i\di_\pp D^+ {\rm\ \ \ and\ \ \ }D^2D_- = -D_- D^2 =
             i\di_\mm D^-\\
   \end{array} \]

In the actual computation of the diagrams, we only need terms with the
 derivatives $(\di_\pp, D_+)$ acting on
  propagators, and so we can effectively drop the $\ln S^\mm$ term in the
 propagator, $G=\ln S^\pp S^\mm$.  We need the following identities:

\begin{eqnarray*}
D_+^z S^\pp = i (\theta-\theta')^+ \equiv i \lp = D_+^{z'} S^\pp \\
  \di_\pp^z G(z,z') = \S{} = -\dizp G(z,z')\ ,  &\quad&
   \Dz G(z,z') = \frac {i\lp}{S^\pp} = \Dzp G(z,z')\\
(\diz)^2 G(z,z') = -\S{2}\ , && (\diz)^3 G(z,z') = \frac{2}{{S^\pp}^3}\\
  \dizp G(z,z') \Lvec{\di^{z'}_\pp} = \S{2} \ ,&&
    \Dz G(z,z') \Lvec{D^{z'}_+} = \frac{i}{S^\pp}
\end{eqnarray*}

   Graphs of the kind shown in Fig.~1(d) are the most complicated.  They
involve integration over two vertices, $u$ and $v$.
We calculate the diagram in Figure~2 as an example:

\let\picnaturalsize=N
\def\picsize{3.0in}
\def\picfilename{fig2.eps}
\ifx\nopictures Y\else{\ifx\epsfloaded Y\else\input epsf \fi
\let\epsfloaded=Y
\centerline{\ifx\picnaturalsize N\epsfxsize \picsize\fi
\epsfbox{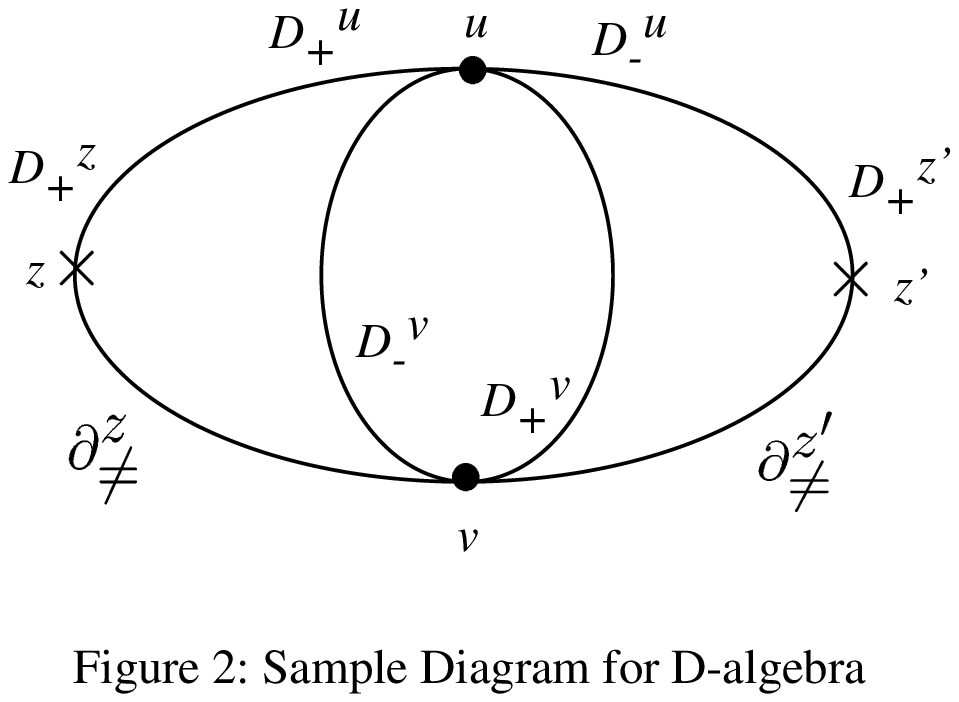}}}\fi

Figure 2 gives:
$$\displaylines{
2 R_{i(jk)l} R^{ijkl} \Int d^4u\, d^4v\ \diz G(z,v)\ D_+^z D_+^u
                    G(z,u)\ D_+^v G(u,v)\ D_-^v G(u,v)\cr
   \hspace{15 em} \times\  \dizp G(v,z')\ D_-^u D_+^{z'} G(u,z')\cr}
$$
  Using (\ref{delta}) to integrate over $u$, and then integrating the $D_-^v$
by
  parts gives:
\begin{eqnarray*}
\lefteqn{
  2 R_{i(jk)l} R^{ijkl} D_+^z D_+^{z'} G(z,z') \Int d^4v\ \diz D_-^v G(z,v) \
      D_+^v G(z',v)\ \dizp G(v,z')\ G(z',v) }&&\\
&=& 2R_{i(jk)l} R^{ijkl} \frac {i}{S^\pp} \Int d^4v\, (-i)\ D_+^v \delta(z-v)\
    D_+^{z'} G(z',v)\ \dizp G(v,z')\ G(z',v)
\end{eqnarray*}
 Now we integrate off the $D_+^v$, and then do the $v$ integral to get:
\begin{eqnarray*}
  \lefteqn{
 2 R_{i(jk)l} R^{ijkl} \frac {i}{S^\pp} \Int d^4v\, i\delta(z-v) (-i)\ \Dz
    D_+^{z'} G(z',v)\ \dizp G(v,z')\ G(z',v)}&&\\
  &=& 2 R_{i(jk)l} R^{ijkl} \frac {i}{S^\pp} i\delta(z-v) (-i)\ \Dz
    D_+^{z'} G(z',z)\ \dizp G(z,z')\ G(z',z)\\
  &=& 2 R_{i(jk)l} R^{ijkl} \frac {i}{{S^\pp}^3} G(z,z')
     \hspace{10 em}     (\times {\rm\ overall\ factors})
\end{eqnarray*}

\end{document}